\def \be {\begin{equation}}
\def \eq {\end{equation}}
\def \bee {\begin{eqnarray}}
\def \eqq {\end{eqnarray}}
\def \nn {\nonumber}
\def \bea {\begin{array}{c}}
\def \eqa {\end{array}}

\def \R {{\bf R}}
\def \C {{\bf C}}
\def \Z {{\bf Z}}
\def \del {\partial}
\def \dels {\partial\kern-.5em / \kern.5em}
\def \As {{A\kern-.5em / \kern.5em}}
\def \Ds {D\kern-.7em / \kern.5em}

\def \a {\alpha}

\def \dag {\dagger}
\def \g {\gamma}

\def \d {\delta}
\def \eps {\epsilon}

\def \s {\sigma}

\def \om {\omega}
\def \Om {\Omega}

\def \th {\theta}

\def \II {I\hspace{-.1em}I\hspace{.1em}}
\def \IIA {\mbox{\II A\hspace{.2em}}}
\def \IIB {\mbox{\II B\hspace{.2em}}}

\def \A {{\cal A}}

\def \Ut {\tilde{U}}
\def \phit {\tilde{\phi}}
\def \phih {\hat{\phi}}
\def \bG {{\bf \Gamma}}

\documentclass[12pt]{article}

\begin{document}
\begin{titlepage}
%\catcode`\@=11
%\catcode`\@=12
%\twocolumn[\hsize\textwidth\columnwidth\hsize\csname%
%@twocolumnfalse\endcsname

%\draft
\begin{center}
%\hfill   UU-HEP/97-07\\
\hfill hep-th/9803166\\

\vskip .5in

\textbf{\large Twisted Bundle On Quantum Torus
and BPS States in Matrix Theory}

\vskip .5in

Pei-Ming Ho
\footnote{on leave from Department of Physics,
University of Utah, Salt Lake City, Utah 84112.}

\vskip .3in

\slshape{Department of Physics, Jadwin Hall, \\
Princeton University, Princeton, NJ 08544}\\
\vskip .2in
\sffamily{pmho@feynman.princeton.edu}

%\maketitle
\end{center}

\vskip .5in

\begin{abstract}

Following the recent work of Connes, Douglas and Schwarz,
we study the M(atrix) model compactified on a torus with
a background of the three-form field. This model is given
by a super Yang-Mills theory on a quantum torus.
To consider twisted gauge field configurations,
we construct twisted $U(n)$ bundles on the quantum torus
as a deformation of its classical counterpart.
By properly taking into account membranes winding around
the light-cone direction, we derive from the M(atrix) model
the BPS spectrum which respects the full
$SL(2,\Z)\times SL(2,\Z)$ U-duality in M theory.

\end{abstract}
%\pacs{PACS numbers: 11.25.-w, 11.25.Mj, 11.25.Sq}%]

\end{titlepage}

%\begin{narrowtext}

\setcounter{footnote}{0}

\section{Introduction} %\hspace{5pt}

In this paper we look into more details of the M(atrix) model
first considered by Connes, Douglas and Schwarz \cite{CDS},
which is a super Yang-Mills theory living on a quantum torus.
It is conjectured that \cite{CDS,DH} this describes
the discrete light-cone quantization (DLCQ) of the M theory
\cite{Suss} compactified on a torus with a nonvanishing
three-form field background.
This idea has been further developed in \cite{HWW,HW,Li,QT}.

The purpose for this work is to derive from this M(atrix) model
the BPS spectrum expected in M theory \cite{CDS} which respects
the full $SL(2,\Z)_C\times SL(2,\Z)_N$ U-duality. Here
$SL(2,\Z)_C$ denotes the group of modular transformations for the
classical torus on which the M theory compactifies, and $SL(2,\Z)_N$
is the non-classical duality group which acts on the three-form
field $C_{123}$. For M theory compactified on $T^3$ with volume
$V$, $SL(2,\Z)_N$ acts on $\tau\equiv C_{123}+iV$ by the linear
fractional transformation
$\tau\rightarrow \frac{a\tau+b}{c\tau+d}$
for any integers $a,b,c,d$ satisfying $ac-bd=1$.
In our case we have one side of the $T^3$ on a ligh-cone direction
so $V=0$ and $\tau=C_{-12}$.
The symmetry of $SL(2,\Z)_C$ is relatively trivial to check in
the M(atrix) model. In this paper we will only focus on the
$SL(2,\Z)_N$ symmetry for the BPS spectrum in M(atrix) model.

However, since we are interested in the BPS configurations
with constant field strength on the quantum torus,
the first question we meet is how to define
twisted (non-trivial) $U(n)$ bundles on a quantum torus.
The general definition of a
bundle on a quantum space is that it is a projective module
of the algebra of functions on the quantum space \cite{Con}.
\footnote{A projective module is a direct summand of a
(finite-dimensional) free module.} In \cite{CR}
the projective module for twisted $U(n)$
bundles over the quantum torus was found and in fact they were used
in \cite{CDS}. Here we will take a naive, classical approach to
construct twisted bundles on the quantum torus from the algebra
of functions on the quantum plane.
Although the results are the same, in this way we show what is the
classical counterpart for this projective module.

The second problem we meet is how to describe in M(atrix) model
membranes winding around one transverse direction as well as the
light-cone direction, because such states are needed to complete
the $SL(2,\Z)_N$ multiplet of BPS states \cite{HV}. We will argue
that in M(atrix) model the winding numbers for such winding states
appear as the quantum numbers parametrizing the Wilson lines for
the electric field background.

In Sec.\ref{comp} we review the Matrix compactification on a torus,
which generically results in the dual space of a quantum torus.
In Sec.\ref{class} we review how sections on twisted bundles over
a classical torus can be constructed for a constant field strength
by imposing twisted boundary conditions. Then we follow the same
steps for the quantum case in Sec.\ref{quant}.
Finally, in Sec.\ref{BPS} we modify the action of
M(atrix) model by inserting appropriate Wilson lines and calculate
the BPS spectrum, which is formally the same as the conjectured BPS
spectrum in \cite{CDS} obtained from M theory arguments,
but we find more independent quantum numbers.

\section{Matrix Compactification On Torus} \label{comp}

In \cite{HWW} and \cite{HW} we gave the general formulation
of Matrix compactification on the quotient space $\R^d/\bG$,
where $\bG$ is a discrete subgroup of Euclidean motions in $\R^d$.
The resulting theory is generically a super Yang-Mills theory (SYM)
on a dual quantum space. The data needed to specify a Matrix
compactification include the action $\Phi$ of $\bG$ on $\R^d$
and a function $\a(g,h)$ with $g,h\in\bG$. We use $\a$ to define
a projective representation of $\bG$ as \cite{HWW,HW}
\be
U_g U_h=e^{i\a(g,h)}U_{gh},
\eq
and then impose the quotient conditions
\be
U_g^{\dag}X^{\mu}U_g=\Phi^{\mu}_g(X).
\eq

Physically inequivalent choices of $\a(g,h)$ are one-to-one
corresponding to elements in the second Hochschild cohomology
$H^2(\bG,U(1))$ defined as follows \cite{HW}.
We call an angular function (defined up to $2\pi$) with
$k$ arguments in $\bG$ a $k$-cochain. The coboundary $\d$ is
an operation which maps a $k$-cochain $\om(g_1,\cdots,g_k)$ to
a $k+1$-cochain given by
\bee
(\d\om)(g_0,\cdots,g_k)&=&\om(g_1,\cdots,g_k)+
\sum_{l=1}^{k}(-1)^l\om(g_0,\cdots,g_{l-1}g_l,\cdots,g_k) \nn\\
&&+(-1)^{k+1}\om(g_0,\cdots,g_{k-1})\eps(g_k),
\eqq
where $\eps(g)=\pm 1$ is the $\Z_2$-grading on $\bG$ corresponding
to the orientifolding. We have $\eps(g_1 g_2)=\eps(g_1)\eps(g_2)$.
It follows that $\d^2=0$.
For orbifolds $\eps(g)=1$ for all $g$.
A cochain annihilated by $\d$ is called a cocycle and $H^2(\bG,U(1))$
is the group of equivalence classes of 2-cocycles defined up to
the coboundary of 1-cochains. It is very important to find out the
corresponding physical degrees of freedom for $H^2(\bG,U(1))$.

Now we focus on the case of the compactification on a 2-torus.
In this case $\bG=\Z^2$ and
for a 2-torus with radii $R_1,R_2$ the quotient conditions are
\be \label{QC}
U_i^{\dag}X_j U_i=X_j+2\pi R_j\d_{ij}, \quad i,j=1,2.
\eq
For simplicity we will consider a straight torus. It is
straightforward to generalize it to a slanted torus.
The algebra of $U_1$ and $U_2$ should satisfy
\be \label{torus}
U_1 U_2=e^{i2\pi\th} U_2 U_1,
\eq
which happens to define the algebra of functions $\A(T^2_{\th})$
on a quantum torus $T^2_\th$.
This $\th$ is the unique parameter for this compactification since
$H^2(\bG,U(1))=U(1)$. It is proposed that \cite{CDS,DH} it
corresponds to the three-form field background $C_{-12}$ for
the DLCQ M theory on a torus in directions $X_1$ and $X_2$:
\be
\th=R C_{-12},
\eq
where $R$ is the radius for light-cone quantization.

For the case $\th=0$ the solutions of $X$ to the quotient conditions
are the covariant derivatives on the dual torus \cite{Tay,GRT}:
$X_i=-i2\pi R_i\del_i+A_i$.
In general, the solutions of $X$ can be viewed as the covariant
derivatives on the dual quantum torus $T^2_{(-\th)}$ \cite{CDS,HWW,HW}.
The algebra of functions on the dual quantum torus $\A(T^2_{(-\th)})$ 
is simply the opposite algebra of $\A(T^2_{\th})$. We denote the
generators of functions on the dual torus by $\Ut_1$ and $\Ut_2$,
with $\Ut_1\Ut_2=e^{-i2\pi\th}\Ut_2\Ut_1$ and $[\Ut_i, U_j]=0$.
Then
\be
X_i=-i2\pi R_i\del_i+A_i(\Ut_1,\Ut_2), \quad i=1,2,
\eq
where $\del_i$ are the derivatives on $T^2_{(-\th)}$:
\bee
&\del_i\Ut_j=\Ut_j(\del_i+i\d_{ij}), \quad
\del_i U_j=U_j(\del_i+i\d_{ij}), \\
&\del_1\del_2=\del_2\del_1.
\eqq

The Matrix compactification specified by (\ref{QC}) and (\ref{torus})
results in the $U(n)$ SYM on the dual quantum torus $T^2_{(-\th)}$.
When the $U(n)$ bundle on the dual torus is untwisted,
the gauge fields $A_i$ are $n\times n$ matrices of functions
on $T^2_{(-\th)}$. A state in the Hilbert space is then an $n$-vector
of functions on $T^2_{(-\th)}$. When the $U(n)$ bundle is twisted, the
Hilbert space is a projective module of $\A(T^2_{(-\th)})$. We will
construct the twisted bundle on the quantum torus by following the naive
classical treatment of $U(n)$ bundles on a torus and view the quantum
case as a deformation.

\section{Twisted Bundles On Classical Torus} \label{class}

Just like we can only talk about functions on a quantum space
rather than points on the quantum space, we can only talk about
sections on a quantum bundle rather than points on the bundle.
Hence we first review the classical construction
for sections on a twisted bundle. Since we are considering $U(n)$
bundles, we can treat sections of the twisted bundle of
fundamental representation as states in a Hilbert space so that
sections of the bundle of adjoint representation as well as
covariant derivatives act as operators on the Hilbert space.

Consider the classical case of a constant curvature with
\be
[D_1, D_2]=-\frac{i}{2\pi}\frac{m}{n}.
\eq
By fixing a gauge, let
\be
D_1=\frac{\del}{\del\s_1}, \quad
D_2=\frac{\del}{\del\s_2}-i\frac{m}{n}\frac{\s_1}{2\pi}.
\eq
The gauge fields are not well-defined functions on $T^2$
but are functions on $\R^2$ satisfying the twisted boundary conditions
\bee
&D_i(\s_1+2\pi,\s_2)=\Om_1(\s_2)D_i(\s_1,\s_2)\Om_1^{\dag}(\s_2), \label{BC1}\\
&D_i(\s_1,\s_2+2\pi)=\Om_2(\s_1)D_i(\s_1,\s_2)\Om_2^{\dag}(\s_1), \label{BC2}
\eqq
where
\be
\Om_1(\s_2)=e^{im\s_2/n}U, \quad \Om_2(\s_1)=V,
\eq
and $U, V$ are $n\times n$ matrices satisfying
\be
UV=e^{-i2\pi m/n}VU.
\eq
The reason for the constant curvature to be quantized by
$m/n$ is that there is a consistency condition on $\Om_i$:
\be
\Om_1(\s_2+2\pi)\Om_2(\s_1)=\Om_2(\s_1+2\pi)\Om_1(\s_2).
\eq

Sections on the twisted bundle of the fundamental representation of
$U(n)$ on a classical 2-torus are $n$-vectors of functions on $\R^2$
satisfying the twisted boundary conditions
\bee
&\phi(\s_1+2\pi,\s_2)=\Om_1(\s_2)\phi(\s_1,\s_2),\\
&\phi(\s_1,\s_2+2\pi)=\Om_2(\s_1)\phi(\s_1,\s_2).
\eqq
In the basis where
\be
U_{kl}=e^{i2\pi km/n}\d_{kl}, \quad V_{kl}=\d_{(k+1)l},
\quad k,l=1,\cdots,n,
\eq
it is not difficult to find the general solution to these
boundary conditions as \cite{GRT}
\be
\phi_k(\s_1,\s_2)=\sum_{s\in\Z}\sum_{j=1}^m
\exp\left\{i\left(\frac{m}{n}(\s_2/2\pi+k+ns)+j\right)\s_1\right\}
\phit_j(\s_2/2\pi+k+ns).
\eq
So a section of the fundamental bundle is specified by
$m$ functions $\phit_j(x)$ with $x\in\R$.

Similarly, sections on the twisted bundle of the adjoint
representation are matrices of functions on $\R^2$ satisfying
the twisted boundary conditions of the same form as (\ref{BC1},\ref{BC2}).
Without loss of generality, we assume that $(n,m)$ are coprime.
Then one can check that two particular sections are given by
\be
Z_1=e^{i\s_1/n}V^{b}, \quad Z_2=e^{i\s_2/n}U^{-b},
\eq
for any two integers $a,b$ satisfying $an-bm=1$.
In fact, for any fixed pair of $a$ and $b$
they generate the algebra of sections on the adjoint bundle.
The actions of $D_i$ and $Z_i$ on the fundamental sections
$\phi_k(\s_1,\s_2)$ induces their actions on $\phit_j(x)$.
These can be read off from the quantum case in the next
section by setting $\th=0$.

\section{Twisted Bundle On Quantum Torus} \label{quant}

Twisted bundles on the quantum torus can be viewed as
a deformation of its classical counterpart.
As in the classical case, sections on a twisted bundle
are not functions well defined on the quantum torus but they are
functions on the quantum plane with twisted boundary conditions.
So we need to extend the algebra of functions on the torus,
which is generated by the two operators $\Ut_1$ and $\Ut_2$,
to the algebra $\A(\C_{(-\th)})$ of functions on the quantum plane,
which is generated by $\s_1$ and $\s_2$ satisfying
\be
[\s_1,\s_2]=i2\pi\th.
\eq
The algebra of functions on $T^2_{(-\th)}$ can be
reailized as a subalgebra of $\A(\C_{(-\th)})$ by
\be
\Ut_i=e^{i\s_i}.
\eq
The derivatives $\del_i$ act on $\s_i$ by
\be
[\del_i,\s_j]=\d_{ij}.
\eq

Now we can repeat what we just did for the classical case.
By fixing a gauge, let
\be
D_1=\frac{\del}{\del\s_1}, \quad
D_2=\frac{\del}{\del\s_2}-if\s_1.
\eq
for the field strength of
\be
[D_1,D_2]=-if.
\eq
The boundary conditions for $D_i$ are (\ref{BC1},\ref{BC2}) with
\be
\Om_1=e^{i\a\s_2}U, \quad \Om_2=V,
\eq
where $\a=\frac{2\pi f}{1+hf}$.
The consistency condition
\be
\Om_1(\s_2+2\pi)\Om_2(\s_1)=\Om_2(\s_1+2\pi)\Om_1(\s_2)
\eq
implies that $\a=m/n$.
Therefore
\be
2\pi f=\frac{m}{n-\th m}.
\eq

The solutions for the boundary conditions for sections
on a fundamental bundle are given by
\be
\phi_k(\s_1,\s_2)=\sum_{s\in\Z}\sum_{j=1}^m
E\left(\frac{m}{n}(\s_2/2\pi+k+ns)+j, i\s_1\right)
\phit_j(\s_2/2\pi+k+ns),
\eq
where the function $E$ is an analogue of the exponential function
but is ``normal ordered'':
\be
E(A,B)\equiv\frac{1}{1-[A,B]}
\sum_{l=0}^{\infty}\frac{1}{l!}A^l B^l,
\eq
where we assumed that $[A,B]$ is proportional to the unity.
It has the following useful properties
\bee
&E(-B,A)E(A,B)=1, \\
&E(A+c,B)=E(A,B)e^{cB}, \\
&E(A,B+c)=e^{cA}E^{A,B}
\eqq
for $c\in\C$.
As in the classical case, the sections on the fundamental bundle
are given by $m$ functions $\phit_j$ on $\R$.

Similarly, the sections on adjoint bundles are generated by
\be \label{Z}
Z_1=e^{i\s_1/(n-\th m)}V^{b}, \quad Z_2=e^{i\s_2/n}U^{-b},
\eq
where $b$ is an integer satisfying $an-bm=1$ for some integer $a$.
Its algebra is
\be \label{ZZ}
Z_1 Z_2=e^{i2\pi\th'}Z_2 Z_1,
\eq
where
\be \label{th-p}
\th'=\frac{b-\th a}{n-\th m}.
\eq
As the $U(n)$ SYM on a classical manifold can also be viewed as
the $U(1)$ SYM on a noncommutative space involving a factor of
$\Z_n$ \cite{HW1}, the twisted bundle with twisting number $m$
on $T^2_{(-\th)}$ can be viewed as the $U(1)$ SYM on a dual
quantum torus $T^2_{\th'}$ \cite{CDS}.
Note that generically $\th'\neq 0$ even when $\th=0$.

Now we view the fundamental sections as the Hilbert space and
let
\be
\phih_j(x)=\phit_j(x-\frac{n}{m}j).
\eq
This Hilbert space is
equivalent to the projective module used in \cite{CDS}.
The action of $D_i$ and $Z_i$ on $\phi_k(\s_1,\s_2)$ induces
their action on $\phih_j(x)$, thus $D_i$ and $Z_i$ can be realized
as operators \cite{CDS}:
\bee
&(D_1\phih)_j(x)=i2\pi fx\phih_j(x),
&(D_2\phih)_j(x)=\frac{1}{2\pi}\frac{\del}{\del x}\phih_j(x), \\
&(Z_1\phih)_j(x)=\phih_{j-a}(x-\frac{1}{m}),
&(Z_2\phih)_j(x)=e^{-i2\pi j/m}e^{\frac{i2\pi x}{n-\th m}}\phih_j(x).
\eqq
It is then easy to compute, e.g.
\be
[D_i, Z_j]=\frac{i}{n-\th m}\d_{ij}Z_j.
\eq

The trace $Tr$ on the adjoint bundle should be a composition of
the trace $tr$ over $n\times n$ matrices and the integration over
$\s_1$ and $\s_2$, i.e., up to normalization,
\be
Tr(f(Z_1,Z_2))=\int d\s_1 d\s_2 tr(f(Z_1,Z_2))
\eq
for a section $f$ of the adjoint bundle. From (\ref{Z}) we see that
\be
Tr(Z_1^{w_1}Z_2^{w_2})=(n-\th m)\d^0_{w_1}\d^0_{w_2},
\eq
where the normalization is chosen such that $Tr(i[D_1,D_2])=m\in\Z$.
While this is the natural normalization for the trace from
the mathematical point of view \cite{CDS}, it also turns out to be the right
choice for the M(atrix) model to respect the $SL(2,\Z)_N$ duality.

\section{BPS Spectrum} \label{BPS}

As a test of the conjecture \cite{CDS,DH} that the SYM on
a quantum torus describes the DLCQ of M theory on a torus
with a three-form field background, we calculate the BPS
spectrum for the M(atrix) model and show that it respects
the $SL(2,\Z)_N$ duality with the quantum torus parameter
$\th$ interpreted as the three-form field background.

The bosonic part of the Lagrangian for the Matrix model in
uncompactified spacetime is
\be
L=Tr\left(\frac{1}{2R}(D_0 X_{\mu})^2+
(2\pi T^M_2)^2\frac{R}{4}[X_{\mu},X_{\nu}]^2\right),
\eq
where $\mu,\nu=1,2,\cdots,9$ and $T^M_2$ is the membrane tension
in M theory. We will set $2\pi T^M_2=1$.
Since we will only consider BPS states without fermions, we will
ignore the fermionic part for simplicity.

We need to modify this action because it misses the description of
the BPS states of membranes winding $m_{i-}$ times around $X_i$
($i=1,2$) and the light-cone direction $X_{-}$.
To see how to do this modification, we note that we can view
this action as the action of Matrix strings \cite{DVV} obtained
from M theory compactified on $X_1$. On the other hand,
the world-sheet action for a string with a $B$ field background
is known to be
\be
L=\frac{1}{2}\del X^{\mu}\bar{\del}X^{\nu}(g_{\mu\nu}+B_{\mu\nu}),
\eq
where $\mu,\nu=0,2,\cdots,9,11$. Let
$X_{\pm}=\frac{1}{\sqrt{2}}(X_0\pm X_{11})$. $X_{-}$ and $X_2$
are compactified on circles with radius $R$ and $R_2$ and
the string lives in a background with $B_{-2}=C_{-12}=\th/R$.
For strings winding $m_{1-}$ times around $X_{-}$ we have
$X_{-}=m_{1-}R\s$. These are understood as membranes winding
$m_{1-}$ times around $X_1$ and $X_{-}$ in M theory.
Thus in the light-cone gauge ($X_{+}=P_{+}\tau$),
the Lagrangian contains a term linear in
$\dot{X}_2$ with the coefficient of $m_{1-}RB_{-2}$.
It follows that in the Hamiltonian the kinetic term
$\frac{1}{2}P_2^2$ is modified to be
$\frac{1}{2}(P_2+m_{1-}RB_{-2})^2$ with $P_2=n_2/R_2$.
Dividing this Hamiltonian by $P_{+}$ and treating it as an
Hamiltonian for $X_2,\cdots,X_9$, with no reference to $X_{11}$,
one finds the corresponding Lagrangian to be
$L=\frac{P_+}{2}\dot{X}_{\mu}^2-\dot{X}_2 m_{1-}RB_{-2}$,
where $\mu=2,\cdots,9$. (If $B_{+-}\neq 0$ there would be an
additional term of $\frac{m_{1-}RB_{+-}}{P_+}$ in $L$.)
This suggests that in the Matrix string Lagrangian we need to
add the term $\frac{\th m_{1-}}{Tr(1)}Tr(\dot{X}_2)$. Similarly,
we can repeat the same argument with the roles of $X_1$,$X_2$
interchanged.

Inserting such appropriate Wilson lines, in the temporal gauge
the action for the M(atrix) model becomes
\be
L=Tr\left(\frac{1}{2R}\dot{X}_i^2+F_{0i}\dot{X}_i
+\frac{1}{2R}\dot{X}_{a}^2+\frac{R}{2}([X_1,X_2]+F_{12})^2
+\frac{R}{2}[X_i,X_a]^2+\frac{R}{4}[X_a,X_b]^2\right)
\eq
with
\be
F_{0i}=\frac{\th m_{i}}{(n-\th m)R_i}, \quad
F_{12}=i2\pi R_1 2\pi R_2\g,
\eq
where $m_i=\eps_{ij}m_{j-}$, $i=1,2$ and $a,b=3,\cdots,9$.
The new term linear in $\dot{X}_i$ appears as a constant electric
field background which was also considered in \cite{Witten,HW2}.
The other modification is that we allowed the background of
a constant magnetic field $\g$ on the dual torus,
which may also be interpreted as the background of a gauge field
coupled to mambranes.

At first it may come as a surprise that the quantum numbers $m_{i-}$
appear in the M(atrix) model not as the eigenvalues of dynamical
variables but as the coefficients of new terms in the action.
However note that these numbers are associated with states
with zero longitudinal momentum, which are supposed to be integrated
out to obtain the light cone theory \cite{DO,HP}.
It is thus natural to expect that
$m_{i-}$ appear as a background or the coefficients of new terms
in the action. Perhaps $\g$ can also have a similar interpretation.

{}From the point of view of SYM theory, the term linear in
$\dot{X}_i$ is a topological term which does not affect the
equations of motion for $X_i$. Imposing periodic boundary conditions
in time on $X_i$, i.e.,
$X_i(\mbox{final})-X_i(\mbox{initial})=2\pi k_i R_i$ for some
integers $k_i$, one finds that the coefficient of $\dot{X}_i$ is
only defined up to an integer over $R_i$ \cite{Witten,HW2}.
Since $\th$ is defined for the quantum torus only up to an integer,
$F_{0i}$ is quantized by an integer $m_{i}$ as above.

It is straightforward to calculate the spectrum for BPS states with
\be
X_i=-i2\pi R_i D_i+p_i t,
\quad X_a=\a_a Z_1^{w_1}Z_2^{w_2} e^{i\om t}
+\a_a^{\dag} Z_2^{-w_2}Z_1^{-w_1} e^{-i\om t},
\eq
where $\om=\frac{2\pi R}{n-\th m}\sqrt{(R_1 w_1)^2+(R_2 w_2)^2}$.
We have
\be
[X_i,Z_1^{w_1}Z_2^{w_2}]=2\pi R_i\frac{w_i}{n-\th m}Z_1^{w_1}Z_2^{w_2}
\eq
and the conjugate momentum for $X_i$ is quantized to be $n_i/R_i$
for the K-K modes.
The BPS spectrum is thus
\bee
H&=&R\left(\frac{1}{2(n-\th m)}\left(\frac{n_i-\th m_i}{R_i}\right)^2
+\frac{1}{2}V^2\frac{(m+(n-\th m)\g)^2}{n-\th m}\right) \nn\\
&&+\frac{2\pi R}{(n-\th m)}\sqrt{(R_1 w_1)^2+(R_2 w_2)^2}, \label{spec}
\eqq
where $V=2\pi R_1 2\pi R_2$.

This spectrum (\ref{spec}) is invariant under the transformation
\bee
&\th\rightarrow -1/\th, \\
&n\rightarrow -m, \quad m\rightarrow n, \\
&n_i\rightarrow -m_i, \quad m_i\rightarrow n_i, \\
&w_i\rightarrow w_i, \\
&\g\rightarrow \th(\th\g-1), \\
&R_i\rightarrow \th^{-2/3}R_i, \quad R\rightarrow \th^{-1/3}R.
\eqq
Together with another symmetry
\bee
&\th\rightarrow \th+1, \\
&n\rightarrow n+m, \quad m\rightarrow m, \\
&n_i\rightarrow n_i+m_{i-}, \quad m_{i-}\rightarrow m_{i-},
\eqq
with everything else unchanged,
they generate the $SL(2,\Z)_N$ symmetry.
The $SL(2,\Z)_N$ transformation on $\th$ is given by
\be
\th\rightarrow \frac{A\th+B}{C\th+D}
\eq
with any integers
$A,B,C,D$ satisfying $AD-BC=1$.
Simultaneously,
\be
\left(\begin{array}{c} n \\ m \end{array}\right)
\rightarrow
\left(\begin{array}{cc} A & B \\ C & D \end{array}\right)
\left(\begin{array}{c} n \\ m \end{array}\right), \quad
\left(\begin{array}{c} a \\ b \end{array}\right)
\rightarrow 
\left(\begin{array}{cc} D & C \\ B & A \end{array}\right)
\left(\begin{array}{c} a \\ b \end{array}\right),
\eq
so that $\th'$ (\ref{th-p}) is invariant.
This implies that the algebra (\ref{ZZ}) of sections on
the bundle of adjoint representation is $SL(2,\Z)_N$-invariant,
although the gauge group is not invariant.
This can be viewed as a hint that the $SL(2,\Z)_N$ symmetry
can be extended to non-BPS states as well.

The interpretation of the quantum numbers is as follows. We have
$n$ as the number of D0-branes, $m$ as the winding number of a
membrane (or the D2-brane number) on the torus of $X_1,X_2$.
The momenta of the K-K modes are $n_i/R_i$ and $m_{i-}$
is the winding number around $X_{-}$ and $X_i$.

In M(atrix) theory the total momentum $P_i=w_i+m\eps_{ij}n_j$
on the dual torus has the same interpretation as the winding
number $nm_{i-}$ \cite{BSS,DVV}. Thus physically we expect to
have the constraint on the winding number $w_i$ given by
$w_i=nm_{i-}-m\eps_{ij}n_j$ as in \cite{CDS}.
In string theory this is understood as the
level matching condition $L_0-\bar{L}_0=0$.
For $\th=\g=m_{i-}=0$ this condition enables one to write
the BPS energy as a complete square
\be
H=\frac{R}{2n}\left(\sqrt{(n_1/R_1)^2+(n_2/R_2)^2}+mV\right)^2,
\eq
which is necessary to interpret the M(atrix) theory on $T^2$
as type \IIB string theory on $S^1$ \cite{Schw}.
However we do not find such a constraint in the M(atrix) model
at finite $N$. The quantum numbers $m_{i-}$ are independent of
the others. This problem of missing the level matching condition
already appears for the M(atrix) model compactified on $S^1$.
In order to obtain type \IIA string theory from the M(atrix) model
on $S^1$ one has to consider the special large $N$ limit and
the level matching condition is achieved only in this limit
\cite{DVV}. It is unclear how to obtain the level matching
condition in M(atrix) theory for a generic background.

It is pointed out in \cite{Li} that to describe the D0-brane
in a B field background in the M(atrix) model limit of \cite{Seib} 
the SYM on a noncommutative torus is not sufficient, rather
the Dirac-Born-Infeld action has to be used. Hence it would be
interesting to see how the SYM results should be modified.

\section*{Acknowledgement}

The author thanks Shyamoli Chaudhuri, Edna Cheung, Michael Douglas,
Eli Hawkins, Danial Kabat, Morton Krogh, Sanjaye Ramgoolam,
Marc Rieffel, Yi-Yen Wu and Yong-Shi Wu for discussions and comments. 
In particular he thanks Michael Douglas for valuable help.
This work was supported in part by NSF grant No. PHY-9601277.

%\end{references}
%\end{narrowtext}


\vskip .8cm

\baselineskip 22pt

\begin{thebibliography}{10}

\itemsep 0pt

\bibitem{CDS}
A. Connes, M. R. Douglas, A. Schwarz,
``Noncommutative Geometry and Matrix Theory:
Compactification on Tori'',
hep-th/9711162.

\bibitem{DH}
M. R. Douglas, C. Hull:
``D-branes and Noncommutative Torus'',
hep-th/9711165.

\bibitem{Suss}
L. Susskind,
`Another Conjecture About M(atrix) Theory'',
hep-th/9704080.

\bibitem{HWW}
P.-M. Ho, Y.-Y. Wu, Y.-S. Wu,
``Towards a Noncommutative Geometric Approach
To Matrix Compactification'',
hep-th/9712201.

\bibitem{HW}
P.-M. Ho, Y.-S. Wu,
``Noncommutative Gauge Theories in Matrix Theory'',
hep-th/9801147.

\bibitem{Li}
M. Li:
``Comments on Supersymmetric Yang-Mills Theory on
a Noncommutative Torus'',
hep-th/9802052.

\bibitem{QT}
N. Nekrasov, A. Schwarz:
``Instantons on Noncommutative $R^4$ and $(2,0)$
Superconformal Six-Dimensional Theory'',
hep-th/9802068;
M. Berkooz:
``Nonlocal Field Theories and the Noncommutative Torus'',
hep-th/9802069;
Y.-K. E. Cheung, M. Krogh:
``Noncommutative Geometry From $0$-Branes in a Background
$B$ Field'',
hep-th/9803031;
F. Ardalan, H. Arfaei, M. M. Sheikh-Jabbari:
``Mixed Branes and M(atrix) Theory on Noncommutative Torus'',
hep-th/9803067.

\bibitem{Con}
A. Connes, ``{\it Noncommutative Geometry}'',
Academic Press, 1994.

\bibitem{CR}
A. Connes,
C.R. Acad. Sci. Paris S'{e}r. A-B 290 (1980) A599;
M. Pimsner and D. Voiculescu,
{\em J. Operator Theory} {\bf 4} 93 (1980);
A. Connes and M. A. Rieffel, ``Yang-Mills for noncommutative two-tori'',
in {\it Operator Algebras and Mathematical Physics} (Iowa City, Iowa, 1985),
pp.237-266, Contemp. Math. Oper. Algebra. Math. Phys. 62, AMS 1987;
M. A. Rieffel, ``Projective Modules Over Higher-dimensional
Non-commutative Tori'', {\em Can. J. Math.} {\bf 40} 257 (1988);
M. A. Rieffel, ``Non-Commutative Tori-A Case Study of Non-Commutative
Differentiable Manifolds'', {\em Contemp. Math.} {\bf 105}, 191-211 (1990).

\bibitem{HV}
F. Hacquebord, H. Verlinde,
``Duality Symmetry of ${\cal N}=4$ Yang-Mills Theory on $T^3$'',
hep-th/9707179,
{\em Nucl. Phys.} {\bf B508}, 609-622 (1997).

\bibitem{Tay}
W. Taylor, IV,
``D-Brane Field Theory On Compact Spaces'',
hep-th/9611042,
{\em Phys. Lett.} {\bf B394}, 283 (1997).

\bibitem{GRT}
O.J. Ganor, S. Ramgoolam and W. Taylor IV,
``Branes, Fluxes and Duality in M(atrix) Theory'',
hep-th/9611202,
{\em Nucl. Phys.} {\bf B492} (1997) 191.

\bibitem{HW1}
P.-M. Ho, Y.-S. Wu,
``D-branes and Noncommutative Geometry",
hep-th/9611233,
{\em Phys. Lett.} {\bf B398}, 251 (1997).

\bibitem{DVV}
R. Dijkgraaf, E. Verlinde, H. Verlinde,
``Matrix String Theory'',
hep-th/9703030,
{\em Nucl. Phys.} {\bf B500}, 43 (1997).

\bibitem{Witten}
E. Witten,
``Bound States of Strings and $p$-Branes'',
hep-th/9510135,
{\em Nucl. Phys.} {\bf B460}, 335 (1996).

\bibitem{HW2}
P.-M. Ho, Y.-S. Wu,
``\IIB/M Duality and Longitudinal Membranes in M(atrix) Theory'',
hep-th/9703016,
{\em Phys. Rev.} {\bf D57}, 2571-2579 (1998).

\bibitem{DO}
M. R. Douglas, H. Ooguri:
``Why Matrix Theory Is Hard'',
hep-th/9710178.

\bibitem{HP}
S. Hellerman, J. Polchinski:
``Compactification in the Lightlike Limit'',
hep-th/9711037.

\bibitem{BSS}
T. Banks, N. Seiberg, S. Shenker:
``Branes From Matrices'',
hept-th/9612157,
{\em Nucl. Phys.} {\bf B490}, 91-106 (1997).

\bibitem{Schw}
J. H. Schwarz:
``An $SL(2,\Z)$ Multiplet of Type \IIB Superstrings'',
hep-th/9508143,
{\em Phys. Lett.} {\bf B360}, 13-18 (1995),
{\em Erratum-ibid.} {\bf B364}, 252 (1995).

\bibitem{Seib}
N. Seiberg:
``Why Is The Matrix Model Correct?'',
hep-th/9710009,
{\em Phys. Rev. Lett.} {\bf 79}, 3577-3580 (1997).

%\bibitem{IKKT}
%N. Ishibashi, H. Kawai, Y. Kitazawa, A. Tsuchiya,
%Nucl. Phys. {\bf 498}, 467 (1997).

\end{thebibliography}
\end{document}